\documentclass[aps,twocolumn,groupedaddress]{revtex4}
\usepackage{epsfig}
\usepackage{graphicx}
\usepackage[T1]{fontenc}
\usepackage{ae}
\usepackage{bbm}
\usepackage[latin1]{inputenc}
\usepackage{amssymb,amsbsy,amsmath}

\begin{document}



\title[Interpolation for spin systems]
{Interpolation between low and high temperatures of the specific heat for spin systems}

\author{Heinz-J\"urgen Schmidt$^1$
\footnote[3]{Correspondence should be addressed to hschmidt@uos.de}
, Andreas Hauser$^2$, Andre Lohmann$^3$ and Johannes Richter$^3$ }

\address{$^1$Universit\"at Osnabr\"uck, Fachbereich Physik,
Barbarastr. 7, D - 49069 Osnabr\"uck, Germany\\
$^2$Abteilung f\"ur Experimentelle Audiologie, Otto-von-Guericke-Universit\"at Magdeburg, D-39016 Magdeburg, Germany \\
$^3$Institut f\"ur Theoretische Physik, Otto-von-Guericke-Universit\"at Magdeburg,
D-39016 Magdeburg, Germany }


\begin{abstract}
The high temperature expansion (HTE) of the specific heat of a spin system fails at low temperatures,
even if it is combined with a Pad\'e approximation. On the other hand we often have information about the low temperature
asymptotics (LTA) of the system. Interpolation methods combine both kind of information, HTE and LTA, in order
to obtain an approximation of the specific heat that holds for the whole temperature range. Here we revisit the entropy method
that has been previously published and propose two variants that better cope with problems of the entropy method for gapped
systems. We compare all three methods applied to the antiferromagnetic Haldane spin-one chain and especially apply the second variant,
called Log Z method, to the cuboctahedron for different spin quantum numbers.
In particular, we demonstrate that the interpolation method is
able to detect an extra low-temperature maximum in the specific heat that
may appear if a separation of two energy scales is present in the considered
system. Finally we illustrate how interpolation also works for classical spin systems.
\end{abstract}

\maketitle

\section{Introduction}\label{sec:I}

The Heisenberg model of interacting localized spins  is an
important canonical model for the theoretical study of quantum many-body
systems \cite{rev1,rev2}. On the other hand, it provides an appropriate description
of the properties of magnetic
insulators \cite{rev1,rev2}. There are numerous theoretical methods to investigate the
model. However, the theoretical description of highly frustrated quantum magnets
is still a very challenging field of solid state theory.
The powerful quantum Monte Carlo methods
 are severely
limited in the presence of frustration due to the infamous ``sign
 problem'' \cite{TrWi05,rev2}. However, for  systems in dimension $D=1$
the density-matrix renormalization group (DMRG) approach \cite{rev2,dmrg}
is very successful to
deal with strongly frustrated quantum spin systems at zero and
finite temperatures.
The situation in $D>1$ is less satisfying. While for $T=0$ several effective
approaches, such as the coupled-cluster method \cite{rev_ccm,ccm2}, the
functional renormalization group approach \cite{rev_fprg,fprg}, the exact
diagonalization \cite{star04,ED40,lauchli2011}, the
extension of the DMRG on $D=2$ \cite{DMRG_D2a,DMRG_D2b}
or the tensor-network approach \cite{tensor}
are available,
suitable methods for finite temperatures and $D>2$ in case of strong
frustration are notoriously rare.

A universal  method to investigate strongly frustrated  quantum magnets is
the high-temperature expansion \cite{domb_green,OHZ06}.
The roots of this method go back to a paper
of W.~ Opechowski published in 1937 \cite{opechowski}.
In
the 1950ies and 1960ies the method was further developed and widely
applied to various Heisenberg systems, see, e.g.,
Refs.~\onlinecite{wood1955,wood1958,wood1967,dalton1969}.
At that time typically the HTE series for specific models could be
calculated until order seven.
In 1990ies an enormous progress calculating the  HTE series for
the Heisenberg model on various lattices could be achieved
by using computer algebraic methods.
Thus, for the frustrated triangular, kagome and
hyperkagome
lattices with only nearest-neighbor exchange bonds for spin quantum number
$s=1/2$ the HTE series is available  up to 14th (16th) order for the triangular
(kagome and hyperkagome) lattice \cite{elstner1993,elstner1994,singh2012}.
However, often we face more complicated systems with more than one exchange
parameter. To determine the HTE series is then more ambitious, since more
complex graphs on the lattice have to be taken into account. As a result,
available series for more complicated models are limited to lower
orders \cite{2DJ1J2,Kapellasite}.
Another difficulty limiting the order of the series arises for higher spin quantum numbers
$s>1/2$, because the HTE series of order $n$ contains $n$th-order polynomials in
$s(s+1)$
which have to be determined for each model separately.
On the other hand,  more complex models as well as higher spin quantum numbers are
relevant for many frustrated magnetic compounds, see, e.g.,
Refs.~\onlinecite{2DJ1J2,Kapellasite,s1,s32}.
Bearing in mind the universal character of the HTE approach it is desirable
to have a general HTE tool to generate the series
for Heisenberg models with arbitrary
exchange patterns and arbitrary spin quantum number $s$.
An early attempt to provide such a tool was published in
Ref.~\onlinecite{SSL01}, where general analytical HTE expressions
up to order three were given.
Very recently some of the present authors have extended this general HTE
scheme up to 10th order \cite{HTE_wir}.
The 10th order scheme is encoded in a simple C++-program and can be
downloaded \cite{url_HTE10} and  freely used by interested researchers.
The `raw' 10th order HTE series provide a good description  of thermodynamic
quantities down to about $1.5 J$, where $J$ is a characteristic exchange energy
of the spin model \cite{HTE_wir}.
The region of validity of the HTE can be significantly extended by  Pad\'e
approximants \cite{baker61} (see also
Refs.~\onlinecite{domb_green} and \onlinecite{OHZ06}). The Pad\'e
approximants are ratios of two
polynomials $[m,n]=P_m(x)/R_n(x)$ of degree $m$ and $n$ and they provide an
analytic continuation of a function $f(x)$ given by a power series.
Such  Pad\'e approximants may yield reasonable data down to about  $0.5 J$.

Having in mind that often we have information on ground-state (and low-energy) properties
obtained by special techniques designed for this purpose, see above, there
remains a gap between the low-temperature description  $T \ll J$ and the Pad\'e-HTE
description at $T \gtrsim 0.5 J$.
This gap is particularly relevant for strongly frustrated magnets, since the
new state of matter in these system appear at temperatures well below $J$.
To bridge this gap, a sophisticated interpolation procedure based on exploiting sum
rules constraining the specific heat was proposed in
Refs.~\onlinecite{BM01} and \onlinecite{MB05},
henceforward called the ``entropy method".
Together with the  general 10th HTE
scheme the interpolation scheme may present a quite universal and powerful
instrument to study the specific heat of frustrated quantum magnets and to
provide model data to compare with experimental results.

To demonstrate this is an aim of the present paper. Another motivation to
reinvestigate the entropy method of Bernu and Misguich \cite{BM01,MB05} and to propose
alternative interpolation schemes consists in the
following:
(i) As for the HTE itself also  for the interpolation  Pad\'e approximants
are used. Often it happens that these approximants exhibit unphysical poles
at temperatures in the region
of interest. Having various interpolation schemes at hand one can
simply exclude such an approximant present in one scheme but absent in
another one.
(ii) For spin systems with an excitation gap $\Delta$ (that is a a quite common
property of systems with a quantum paramagnetic ground state)
the approach of Bernu and Misguich is based on the special ansatz
$c(T)\sim A\; T^{-2}e^{-\Delta /T}$
for the low-temperature behavior of the specific heat which might be not the
correct one for the considered gapped spin system. One example is the
Haldane spin-one antiferromagnetic spin chain, where this special ansatz
has to be replaced by $c(T)\sim A\; T^{-3/2}e^{-\Delta /T}$.

The entropy method is based on the idea not to interpolate the specific heat $c(T)$
directly but rather the entropy $s$ considered as a function $s(e)$ of the energy $e$
utilizing the behavior of $s(e)$ for the two limits of low and high temperatures.
The first modification of this method consists of representing the graph of $s(e)$ in
parametric form as $s(\beta)$ and $e(\beta)$ where $\beta=1/T$ and to interpolate both parametrizations
separately. This gives the ``modified entropy method" that avoids the mentioned problem
with the LTA ansatz.

Another alternative, still in the spirit of the entropy method, is based on the interpolation
of $\ell(\beta)\equiv \log Z(\beta)$  where $Z$ denotes the partition function or rather its thermodynamic limit.
The present paper focuses on this method, called the ``Log Z method". It works for rather general LTA
including finite spin systems and classical ones.

The paper is organized as follows. First we sketch the basic ideas underlying the various interpolation schemes.
In section \ref{sec:P} we explain  what we will call ``pure interpolation" using a toy example.
The specific heat satisfies two integral constraints or ``sum rules". Interpolation with constraints, to be considered in section \ref{sec:C},
proceeds by pure interpolation of another thermodynamic function and by deriving $c(T)$ from this function.
This function will be $s(e)$ for the entropy method, subsection \ref{sec:CE}, or the functions $s(\beta)$ and $e(\beta)$ for
the modified entropy method, subsection \ref{sec:CM}, or the function $\ell(\beta)$ for the Log Z method, subsection \ref{sec:CL}.

The next section \ref{sec:A} contains tests and applications. First, in subsection \ref{sec:AH}, we revisit the example
already considered in \cite{BM01}, the $s=1$ Haldane chain, and show that all three methods yield comparable results
despite the different forms of the LTA. The second subsection \ref{sec:AC} is devoted to the cuboctahedron and the Log Z method.
This is a finite spin system that can be numerically exactly solved for
$s=1/2$, $s=1$, and $s=3/2$ and thus serves as a test for the
considered interpolation scheme. In particular, we want to find out whether
an additional low-temperature maximum present in $c(T)$ for $s=1/2$ and  $s=1$ can be
detected by the interpolation scheme.
For higher $s$, where no exact data are
available,  the Log Z method yields predictions about the form of $c(T)$ for
low and intermediate
temperatures that are hardly available by other methods.
Finally, in section \ref{sec:CLA}, we extend the range of applications
to classical spin systems where a modified second integral constraint has to be considered. We conjecture that the LTA
of the specific heat has a similar form as for gapped quantum systems and verify this conjecture in subsection \ref{sec:CLAT}
for the special case of the classical equilateral spin triangle that can be solved analytically. In the appendix \ref{sec:LTA} we show
how to derive the LTA of the classical spin triangle without using the analytical solution.

\section{Pure interpolation}\label{sec:P}
To illustrate the basic ideas of the interpolation
we consider a toy example. Let the error function $\mbox{erf}(\beta)$ be the unknown thermodynamical
function, where $\beta$ is the inverse temperature,
and assume that we know a finite number of terms of its high temperature expansion (HTE)
\begin{equation}\label{P1}
\mbox{erf}(\beta) = \frac{2}{\sqrt{\pi}} \beta -\frac{2}{3\sqrt{\pi}} \beta^3 +\ldots
\;,
\end{equation}
as well as its low-temperature asymptotic (LTA)
\begin{equation}\label{P2}
\mbox{erf}(\beta) \sim \mbox{erf}_\infty(\beta)\equiv 1 -\frac{e^{-\beta^2}}{\sqrt{\pi}\beta}\;,
\quad \beta\rightarrow\infty
\;.
\end{equation}
The LTA of a thermodynamic function will always be indicated by the subscript $\infty$ reminding that
this is the case of $\beta\rightarrow\infty$.
The basic idea of interpolation is to multiply the LTA by a Pad\'e approximant, e.~g.,
\begin{equation}\label{P3}
\mbox{erf}_{\mbox{\scriptsize int}}(\beta) = \mbox{erf}_\infty(\beta)\frac{a_0+a_1\beta+a_2\beta^2}{1+a_3 \beta+a_2 \beta^2}
\;.
\end{equation}
We have chosen the same coefficient $a_2$ for the leading coefficient of the numerator and the denominator of the
Pad\'e approximant in order to guarantee  $\mbox{erf}_{\mbox{\scriptsize int}}(\beta)\rightarrow \mbox{erf}_\infty(\beta)$
for $\beta\rightarrow\infty$.
Further we have to choose the coefficients $a_0,\ldots,a_3$ such that the HTE of $\mbox{erf}_{\mbox{\scriptsize int}}(\beta)$ coincides
with the known HTE of $\mbox{erf}(\beta)$ up to $3$rd order. Here we face a problem that is typical for the
interpolation of thermodynamical functions, namely that $\mbox{erf}_{\mbox{\scriptsize int}}(\beta)$ is not
analytical at $\beta=0$ due to the factor $\frac{1}{\beta}$ in (\ref{P2}). Hence one could not calculate its HTE
as required by the method.

There are different ways to cope with this problem.
The first one is to modify the Pad\'e approximant such that $\mbox{erf}_{\mbox{\scriptsize int}}(\beta)$ becomes analytical and to write
\begin{equation}\label{P4}
\mbox{erf}_{\mbox{\scriptsize int}}(\beta) = \mbox{erf}_\infty(\beta)\frac{a_1\beta+a_2\beta^2}{1+a_3 \beta+a_2 \beta^2}
\;.
\end{equation}
The Pad\'e approximant has now a series expansion that starts with $a_1\beta$ and cancels the disturbing factor $\frac{1}{\beta}$.
In our toy example we would then have to determine four unknown coefficients and need the HTE of $\mbox{erf}$
up to third order.

Another way out especially employed in \cite{BM01} is the use of auxiliary functions. In our case the function
\begin{equation}\label{P5}
G(\beta)\equiv (1- \mbox{erf}_\infty(\beta))^{-1}=\sqrt{\pi}\,\beta\,\exp(\beta^2)
\;,
\end{equation}
is analytical at $\beta=0$ and the corresponding interpolation ansatz reads
\begin{equation}\label{P6}
G_{\mbox{\scriptsize int}}(\beta) = \sqrt{\pi}\,\beta\,\exp(\beta^2)\frac{a_0+a_1\beta+a_2\beta^2}{1+ a_3\beta+a_2 \beta^2}
\;.
\end{equation}
The rest is analogous to the above procedure; we choose $a_0,a_1,a_2$ such that the HTE of
$G_{\mbox{\scriptsize int}}(\beta)$ coincides with the known HTE of $(1- \mbox{erf}(\beta))^{-1}$ up to second order.
Of course, one has to solve the result for $\mbox{erf}_{\mbox{\scriptsize int}}$.
It is not necessary to provide the further details of the toy example. It turns out that already the
second order of HTE is sufficient to obtain a reasonably good  interpolation of the $\mbox{erf}$ function.
But this is obviously due to the smooth behavior of $\mbox{erf}$ between the low- and the high-temperature
regime. The specific heat has one or even two maxima for intermediate temperatures and thus
represents a greater challenge for interpolation.

By the second method we can also treat the case where some constants in the LTA are unknown.
Assume, for example, that we only know that the LTA of  $\mbox{erf}$ is of the form
\begin{equation}\label{P7}
\mbox{erf}(\beta) \sim \mbox{erf}_\infty(\beta)\equiv 1 -A \frac{e^{-\beta^2}}{\beta}\;,
\quad \beta\rightarrow\infty
\;.
\end{equation}
In this case we would replace (\ref{P6}) by
\begin{equation}\label{P8}
G_{\mbox{\scriptsize int}}(\beta) = \beta\,\exp(\beta^2)\frac{a_0+a_1\beta+a_2\beta^2}{1+ a_3\beta+a_4 \beta^2}
\;,
\end{equation}
and determine $A$ by the ration $a_2/a_4$ as a by-product of the Pad\'e approximation.
In what follows we refer to the above illustrated scheme as {\it pure
interpolation}.

\section{Interpolation with constraints}\label{sec:C}

As mentioned in the Introduction the task of interpolating the specific heat
becomes more difficult if we try to allow for the two integral constraints
\begin{eqnarray}\label{C1a}
\int_0^\infty c(T) dT &=& \int_{T=0}^{T=\infty} de = e_\infty-e_0 = |e_0|,\\
\label{C1b}
\int_0^\infty \frac{c(T)}{T} dT &=&\int_{T=0}^{T=\infty} ds = s_\infty-s_0 =
\log(2s+1),
\end{eqnarray}
where we have assumed that the entropy $s_0$ of the ground state vanishes.
On the other hand the interpolation becomes more accurate by taking into account
this information.
The basic idea of \cite{BM01} is to reduce the task of interpolation with constraints
to a pure interpolation of the function $s=s(e)$. Hence this method has been dubbed
the ``entropy method".
We will shortly recapitulate it following the ideas of Bernu
and Misguich \cite{BM01,MB05}  in the next subsection and then sketch two extensions,
the ``modified entropy method" and the `` log Z method".

\subsection{The entropy method}\label{sec:CE}

\begin{figure}
\begin{center}
\includegraphics[clip=on,width=70mm,angle=0]{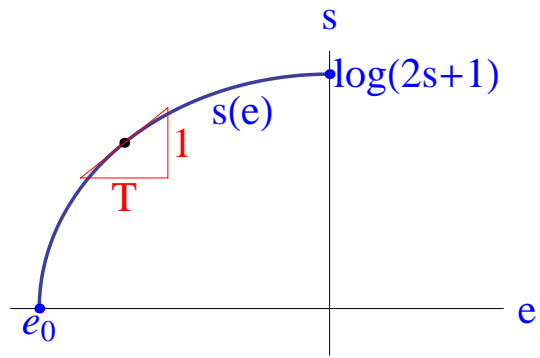}
\end{center}
\caption{Typical graph of the function $s=s(e)$. The slope is the inverse temperature
$\beta=\frac{1}{T}$ due to the Clausius formula $ds =\frac{de}{T}$.
}
\label{fig_entropy}
\end{figure}

For a spin system the graph of the function $s=s(e)$ typically has the form shown in Fig.~\ref{fig_entropy}.
It is limited by the points $(s=0, e=e_0)$ and $(s=\log(2s+1),e=0)$ corresponding to $T=0$ and $T=\infty$.
Since
\begin{equation}\label{CE1}
\frac{d\,s}{d\,e} = \beta = \frac{1}{T}
\;,
\end{equation}
the slope of the graph vanishes at $T=\infty$ and diverges at $T=0$.
The specific heat $c=c(e)$ can be obtained by
\begin{eqnarray}\label{CE2a}
\frac{1}{c(e)} &=& \frac{d\,T}{d\,e} = \frac{d}{d\,e} \left(\frac{d\,s}{d\,e}\right)^{-1} = -\frac{s''(e)}{s'(e)^2}\;,\\
\label{CE2b}
c(e)&=&  -\frac{s'(e)^2}{s''(e)}
\;.
\end{eqnarray}
The entropy method assumes that we know the HTE and the LTA of
$s(e)$. Whereas the HTE can be derived from that of $c(\beta)$
without problems, the derivation of an explicit LTA may be
problematic as we will see below. But if we assume the HTE and LTA
of $s(e)$ as given we may perform a pure interpolation by one of the
methods sketched in section \ref{sec:P} and obtain some
approximation $s_{\mbox{\scriptsize int}}(e)$ of $s(e)$. From this
we obtain the corresponding $c_{\mbox{\scriptsize int}}(e)$ by
(\ref{CE2b}) and $\beta_{\mbox{\scriptsize int}}(e)$ by (\ref{CE1}).
The plot of the final result $c=c_{\mbox{\scriptsize int}}(\beta)$ can be obtained
in parametric form from these results or by inserting the numerically obtained inversion of $\beta_{\mbox{\scriptsize int}}(e)$.
It necessarily satisfies the integral constraints (\ref{C1a}), (\ref{C1b})
and has the correct behavior for low and high temperatures.

In the applications of the interpolation method we find three cases of the LTA of the specific heat, namely
\begin{eqnarray}\label{CE3a}
c(T)&\sim &A\; T^{\alpha}\;,\\
\label{CE3b}
c(T)&\sim &A\; T^{\alpha}e^{-\Delta /T}\;,\\
\label{CE3c}
c(T)&\sim & T^{-2}\sum_{\nu=1}^n d_\nu\delta_\nu^2 e^{-\delta_\nu /T}.
\end{eqnarray}
The case (\ref{CE3a}) occurs for gapless systems, (\ref{CE3b}) for systems with an energy gap $\Delta$ and
(\ref{CE3c}) for finite systems. The latter case will be considered in more details in section \ref{sec:CL}
where also the notation will be explained.
The entropy method applies to the cases (\ref{CE3a}) and (\ref{CE3b}), where
the latter case has to be restricted  to $\alpha=-2$.
Moreover, the case (\ref{CE3c}) with $n=1$ could be treated by this method.
In the other cases
one has the problem to find an explicit form of the LTA of $s(e)$. This is a motivation to extend the
entropy method in order to cover all cases.

\subsection{The modified entropy method}\label{sec:CM}

The crucial object of the entropy method is the graph of $s=s(e)$. It can also be represented
in parametric form by two functions $e=e(\beta)$ and $s=s(\beta)$. Here we have chosen the inverse
temperature $\beta$ as the natural parameter; but sometimes it can be more convenient to choose
other parameters such as $\sqrt{\beta}$ and to adapt the modified entropy method to this choice.
We assume that for both functions $e=e(\beta)$ and $s=s(\beta)$ a pure interpolation is possible.
For example, in the case (\ref{CE3b}) we could obtain an LTA of $e=e(\beta)$ and $s=s(\beta)$ in the following way:
\begin{eqnarray}\nonumber
e_\infty(\beta)&=&e_0-\int_\infty^\beta c_\infty(x)\frac{dx}{x^2} =e_0-A\int_\infty^\beta x^{-\alpha} e^{-\Delta x}\frac{dx}{x^2}\\
\nonumber
&=& e_0+A\left(
\frac{\beta^{-2-\alpha}}{\Delta}e^{-\Delta\beta}+\int_\infty^\beta \frac{2+\alpha}{\Delta}x^{-3-\alpha} e^{-\Delta x}\,dx
\right)\\
\label{CM1a}
&\sim &e_0+A\;
\frac{\beta^{-2-\alpha}}{\Delta}e^{-\Delta\beta}\;,\\
\nonumber
s_\infty(\beta)&=&-\int_\infty^\beta c_\infty(x)\frac{dx}{x} =-A\int_\infty^\beta x^{-\alpha} e^{-\Delta x}\frac{dx}{x}\\
\nonumber
&=& A\left(
\frac{\beta^{-1-\alpha}}{\Delta}e^{-\Delta\beta}+\int_\infty^\beta \frac{1+\alpha}{\Delta}x^{-2-\alpha} e^{-\Delta x}\,dx
\right)\\
\label{CM1b}
&\sim &A\;
\frac{\beta^{-1-\alpha}}{\Delta}e^{-\Delta\beta}\;.
\end{eqnarray}
We have obtained this result by partial integration and taking only the terms with the highest power
of $\beta$. It is clear from (\ref{CM1a}) that $e_\infty(\beta)$ cannot be solved for $\beta$ except for
$\alpha=-2$, and hence the entropy method cannot be applied to gapped systems in the general case.
For the modified entropy method it is not necessary to solve $e_\infty(\beta)$ for $\beta$
since we apply the pure interpolation procedure directly to $e(\beta)$ and $s(\beta)$. Hence this method can be also applied
to gapped systems with $\alpha\neq-2$.

From the approximated graph given by the parametric representation
$e_{\mbox{\scriptsize int}}(\beta)$ and $s_{\mbox{\scriptsize int}}(\beta)$ we can obtain the specific heat
$c(\beta)$ by
\begin{equation}\label{CM2}
c(\beta)= -\frac{\dot{s}^2\dot{e}}{\ddot{s}\dot{e}-\dot{s}\ddot{e}}
\;.
\end{equation}
Here the dot indicates the derivative w.~r.~t.~$\beta$ and the result
easily follows from (\ref{CE2b}) by the chain rule.

Applications of the modified entropy method will be presented in section \ref{sec:AH}.
As a drawback of this method we note that it results in two different temperature
concepts, namely $T_1=\frac{1}{\beta}$ and $T_2=\frac{\dot{s}}{\dot{e}}$ that only
coincide for the correct graph. For the approximated graph there will be a slight difference
between $T_1$ and $T_2$, especially at low temperatures. It remains open which temperature
one should choose for the final result of $c(T)$; $T_1$ gives the best results for the
behavior at high and low temperatures and $T_2$ exactly satisfies the integral constraints
(\ref{C1a}),(\ref{C1b}).

\subsection{The Log Z method}\label{sec:CL}

This method is based on the well-known fact that both functions, $s(\beta)$ and $e(\beta)$,
can be derived from a single thermodynamic function, namely $\log Z= \log \mbox{Tr}\left( e^{-\beta H}\right)$ or, more precisely,
its thermodynamic limit
\begin{equation}\label{CL1}
\ell(\beta)\equiv \lim_{N\rightarrow\infty}\frac{1}{N} \log Z_N
\;.
\end{equation}
In fact,
\begin{eqnarray}\label{CL2a}
e(\beta)&=&-\frac{d}{d\beta}\ell(\beta)\;,\\
\label{CL2b}
c(\beta)&=&\beta^2\frac{d^2}{d\beta^2}\ell(\beta)\;,\\
\label{CL2c}
f(\beta)&=&-\frac{1}{\beta}\ell(\beta)\;,\\
\label{CL2d}
s(\beta)&=&\beta\left(e(\beta)-f(\beta)\right)\;.
\end{eqnarray}
If we can determine the correct behavior of $\ell(\beta)$ for low and high
temperatures,
we can perform a pure interpolation and obtain $\ell_{\mbox{\scriptsize int}}(\beta)$.
Then the functions $e_{\mbox{\scriptsize int}}(\beta)$ and $s_{\mbox{\scriptsize int}}(\beta)$ defined by
(\ref{CL2a}) and (\ref{CL2d}) will automatically inherit the correct low and high temperature behavior.
It follows from Eqs.~(\ref{CL2a}) and (\ref{CL2d})
that $c_{\mbox{\scriptsize int}}(\beta)$ defined by (\ref{CL2b}) satisfies the integral constraints
(\ref{C1a}),(\ref{C1b}). Moreover,  $c_{\mbox{\scriptsize
int}}(\beta)$ also has the correct LTA and HTE of the considered order.

Let $c(\beta)=\sum_{n=0}^\infty  d_n \beta^n$ be the HTE of the specific heat and
$\ell(\beta)=\sum_{n=0}^\infty  a_n \beta^n$ that of Log Z, then (\ref{CL2b}) implies
that $a_n=\frac{d_n}{n(n-1)}$ for all $n=2,3,\ldots$. Further,
$a_0=\lim_{N\rightarrow\infty}\frac{1}{N}\log \mbox{Tr}\,{\mathbbm 1}= \log (2s+1)$
by the definition (\ref{CL1}), and $a_1=\mbox{Tr}H = 0$.  Hence, from a
known HTE series of the specific heat $c$ up to order $n_{\mbox{\scriptsize max}}$ follows  the HTE of
$\ell(\beta)$ up to the same order.

For the LTA we first consider a finite system with energy eigenvalues $e_0,e_1,e_2,\ldots$ and the corresponding
degeneracies $d_0=1,d_1,d_2,\ldots$. Of course, the limit $N\rightarrow \infty$ is ignored for finite systems.
Set $\delta_\nu \equiv e_\nu-e_0$ for $\nu=1,2,\ldots$.
We thus have
\begin{eqnarray}\label{CL3a}
Z&=& e^{-\beta e_0}+d_1\, e^{-\beta e_1}+d_2\, e^{-\beta e_2}+\ldots\\
\label{CL3b}
&=& e^{-\beta e_0}(1+d_1\, e^{-\beta \delta_1}+d_2\, e^{-\beta \delta_2}+\ldots)
\;,
\end{eqnarray}
and hence
\begin{eqnarray}\nonumber
\log Z&=& -\beta e_0 +\log \left(1+d_1\, e^{-\beta \delta_1}+d_2\, e^{-\beta \delta_2}+\ldots\right)\\
&&\label{CL4a}\\
&\sim& -\beta e_0 + \sum_{\nu=1}^m d_\nu e^{-\beta \delta_\nu}
\;\mbox{for  } \beta\rightarrow\infty \;,
\end{eqnarray}
where we have truncated the summation at $\nu=m$ such that $\delta_m<2 \delta_1$. Otherwise we should
have included the second term of the Taylor series $\log(1+x) = x-\frac{x^2}{2}+\ldots$. This result
also yields the LTA of the specific heat in (\ref{CE3c}) by means of (\ref{CL2b}). Note that the term
$ -\beta e_0 $ in (\ref{CL4a}) has the effect that the LTA of $e(\beta)$ as well as of
$f(\beta)$ starts with the term $e_0$, as it is expected.

Now we will assume that also for an infinite gapped spin system the LTA of $\ell(\beta)$ is of the form
\begin{equation}\label{CL5}
\ell_\infty(\beta)=-\beta e_0 + \frac{A}{\Delta^2}\beta^{\alpha-2}e^{-\Delta\beta}
\;,
\end{equation}
where $e_0$ has to be redefined as the ground state energy per spin in the thermodynamic limit.
The notation is chosen such that (\ref{CE3b}) is obtained as the leading term of the LTA
of $c(\beta)$.

In the applications to lattice spin systems it may happen that only $\Delta$ is known
but $A$ is unknown or that both, $\Delta$ and $A$, are unknown. Hence it is important to note that,
like the entropy method, also the Log Z method is principally able to obtain estimates of these data via interpolation
in the way indicated at the end of section \ref{sec:P}.

This finishes the general explication of the log Z method. Further details
will be given in the following sections including applications to concrete systems.

\section{Tests and applications}\label{sec:A}

\begin{figure}
\begin{center}
\includegraphics[clip=on,width=70mm,angle=0]{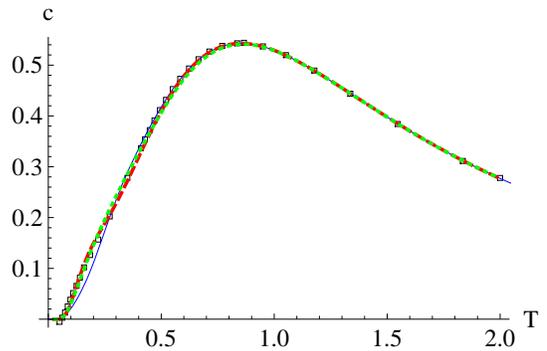}
\end{center}
\caption{Comparison of the curves $c$ vs. $T$ for the $s=1$ Haldane chain
determined by four different methods. The green dotted curve is obtained by the entropy method
\cite{BM01}, the blue one by the modified entropy method, the red dashed one by the Log Z method, and the curve indicated by small
black squares by DMRG methods, see \cite{X98}, Fig.~4.
 } \label{figComp1}
\end{figure}

\begin{figure}
\begin{center}
\includegraphics[clip=on,width=70mm,angle=0]{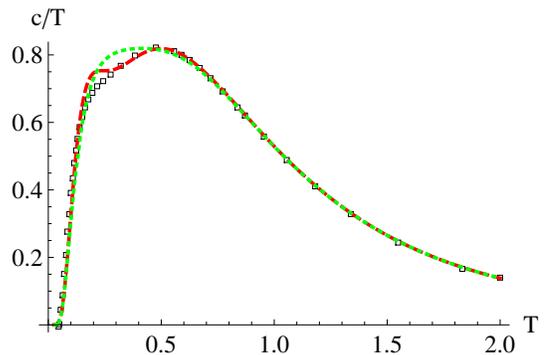}
\end{center}
\caption{Comparison of the curves $c/T$ vs. $T$ for the $s=1$ Haldane chain
determined by the following three different methods. The green dotted curve is obtained by the entropy method
\cite{BM01}, the red one by the Log Z method, and the curve indicated by small
black squares by DMRG methods, see \cite{X98}, Fig.~4.
} \label{figComp2}
\end{figure}

\subsection{The $s=1$ Haldane chain}\label{sec:AH}
The $s=1$ antiferromagnetic Heisenberg spin chain (``Haldane chain") is an example of a gapped system where
the LTA of the specific heat is known \cite{JG94} to be of the form
\begin{equation}\label{AH1}
c_\infty(T)=\frac{\Delta^{5/2}}{\sqrt{2\pi}}T^{-3/2} e^{-\Delta/T}
\;.
\end{equation}
The values of the ground state energy $e_0=-1.401484038971(4)$ and the gap $\Delta=0.41050(2)$
have been determined by DMRG calculations \cite{WH93}. Also the specific heat has been calculated
by these techniques that are especially suited for one-dimensional systems \cite{X98}. Hence this system can be used
as a kind of test for interpolation methods.

Bernu and Misguich \cite{BM01} have applied the entropy method to the Haldane chain although the exponent in (\ref{AH1})
is $\alpha=-3/2$ by simply setting $\alpha=-2$. They used unpublished HTE data up to the order 20 of
Elstner, Jolicoeur, and Golinelli and an auxiliary function
\begin{equation}\label{AH2}
G(e)=(e-e_0)\frac{d}{de}\frac{s(e)}{e-e_0}
\;.
\end{equation}
By this they determined $\Delta$ by extrapolation w.~r.~t.~the order $n$ approximately to $\Delta\approx 0.40$.

The modified entropy method and the Log Z method can be directly applied to the Haldane chain using the correct
exponent $\alpha=-3/2$ of the LTA of $c(T)$. We have done this for both methods by using the same HTE data as
\cite{BM01}.
Since we are able to use the true exponent $\alpha=-3/2$, we
use the full information about the gap and the prefactor of the LTA of the specific heat given
above.
We will explain some details of the procedure
only for the Log Z method.

The LTA (\ref{AH1}) is not analytic in $\beta$ due to the factor $\beta^{3/2}$. Instead of using an auxiliary function we
will make a transformation to the independent variable $\gamma\equiv \sqrt{\beta}$. Hence our ansatz for the interpolation
of $\ell(\gamma)$ assumes the form
\begin{eqnarray}\nonumber
\ell_{\mbox{\scriptsize int}}(\gamma)&=&-\gamma^2 e_0 +\sqrt{\frac{\Delta}{2\pi}}e^{-\Delta\gamma^2} \gamma^{-1}\,\times\\
\label{AH3}
&& \frac{a_1\gamma +\ldots +a_{2N} \gamma^{2N}}{1+a_{2N+1} \gamma+\ldots+a_{2N} \gamma^{2N}}
\;.
\end{eqnarray}
Note that we have chosen the coefficients of the Pad\'e function such that its numerator contains the factor $\gamma$
that cancels the  $\gamma^{-1}$ of $\ell_\infty(\gamma)$.
Moreover we can choose $2N=40$ since the HTE of $\ell(\gamma)$ is used up to the order $40$ w.~r.~t.~$\gamma$.
The final result has to be transformed back to the variable $\beta$.

We show the results of the three interpolation methods in Fig.~\ref{figComp1} together with DMRG results obtained by \cite{X98}.
The differences are rather small and only visible for
low temperatures. The deviation at low temperatures of the curve corresponding to the modified entropy method from the other curves
is probably due to the fact that it only uses HTE results up to $17$th order because of problems with spurious poles.

In order to better demonstrate the differences at low temperatures we have also plotted the results for $c/T$ vs. $T$ obtained by
the entropy method \cite{BM01}, the Log Z method and the DMRG method
\cite{X98}, see  Fig.~\ref{figComp2}. The results according to the modified
entropy method have not been included because of the larger deviations at low temperatures mentioned above.
It appears that all three methods coincide in the large,
although the two latter methods indicate the occurrence of a shoulder
at $T\approx 0.2$, most markedly by the Log Z method. However, these findings
do not favor one of the three interpolation methods and are rather suited for a positive test of consistency.

\subsection{The cuboctahedron}\label{sec:AC}

\begin{figure}
\begin{center}
\includegraphics[clip=on,width=70mm,angle=0]{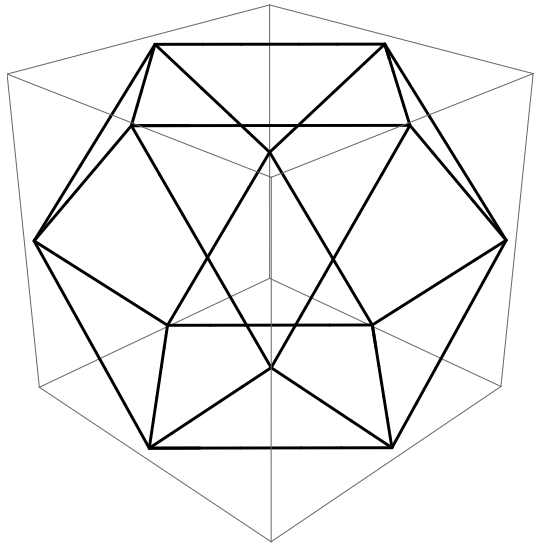}
\end{center}
\caption{The cuboctahedron is an Archimedean solid that results from
joining the $12$ midpoints of the edges of a cube. } \label{figC0}
\end{figure}

Note first that in this section $\ell$ stands for the
full Log Z (i.e. without division by $N$).
The cuboctahedron is an Archimedean solid that results from joining
the $12$ midpoints of the edges of a cube, see Fig.~\ref{figC0}.
It is an example of a
finite spin system with $N=12$, where usually the spins are placed at the vertices and
coupled to their nearest neighbors by the Heisenberg interaction model.
For several reasons the investigation of the resulting Heisenberg model is
an interesting problem. First of all, it is a paradigmatic model to study
frustration
effects
\cite{schmidt2005,MCE2007,schnalle2009,honecker2009,dalton2010,Entel2011,Strecka2016}. Because the
cuboctahedron is built by corner sharing triangles it can be considered as
a finite-size relative
of the celebrated  kagome lattice, see, e.g.,
Refs.~\onlinecite{ccm2,lauchli2011,DMRG_D2a,DMRG_D2b,tensor} and references therein.
For
spin quantum numbers $s=1/2,1,\ldots,3$ the low-lying eigenstates of the quantum
Heisenberg model were calculated by Lanczos exact
diagonalization \cite{schmidt2005}. Here we add corresponding data for
$s=7/2$. For lower values of the spin quantum number, $s=1/2,1,3/2$,
the complete spectrum of eigenvalues can be calculated, i.e., numerically exact
data for thermodynamic quantities are
available \cite{schmidt2005,schnalle2009}.
Indeed, the low-temperature thermodynamics of the quantum model exhibits some interesting
features, such as an extra low-temperature maximum in the specific heat
that indicates a separation of energy scales.

Hence, the cuboctahedron may serve as a nontrivial model system to test the above illustrated interpolation method.
In particular, the question arises whether
the additional low-temperature maximum of $c(T)$ can be detected by the
interpolation approximation.
Moreover, this system will serve to illustrate the flexibility of the Log Z interpolation
method w.~r.~t.~the use of information about the low-lying part of the spectrum.
Another reason to investigate the Heisenberg model on the cuboctahedron is its relation to
experiments on magnetic molecules \cite{cuboc_Exp1,cuboc_Exp2}.

\begin{figure}[ht]
\hspace*{-1.5cm}
\vspace*{1.5cm}
\includegraphics[clip=on,width=110mm,angle=0]{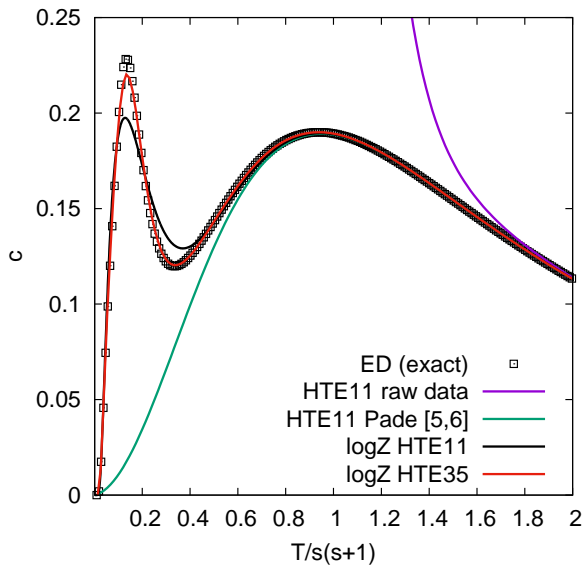}
\caption{
Specific heat $c(T)$ of the cuboctahedron for $s=1/2$, calculated by the Log Z method compared to the exact function
(ED, open squares).
The black solid line shows the  result using one excitation and an HTE of order $n=11$,
the  red solid line shows  analogously data for the $4$ lowest excitations and
$n=35$.
For comparison we also show the `raw' HTE of order $n=11$ (blue curve) as well
as the [5,6] Pad\'e approximant (green curve).
} \label{figC1}
\end{figure}

To be more specific, we illustrate the Log Z method using the values and the degeneracies of the first
$k$
eigenvalues of the Hamiltonian as well as the HTE of the specific heat of order
$n=11$, i.e. we can use 12 HTE coefficients of $\ell(\beta)$ as input for
the interpolation, cf. section \ref{sec:CL}.
This use of HTE data of order $n=11$ is based on unpublished results that exceed the
published ones \cite{HTE_wir},\cite{url_HTE10} by one order.

We write
\begin{eqnarray}\nonumber
\ell_{\mbox{\scriptsize int}}(\beta)&=&\sum_{i=1}^{k}d_i\,\exp(-\beta\,\delta_i)\,
\frac{\sum_{n=0}^6 a_n \beta^n}{1+\sum_{n=1}^5 a_{n+6}\beta^n\,+a_6\beta^6}\\
\label{AC1}
&&-\beta\;e_0
\;.
\end{eqnarray}
The Pad\'e coefficients $a_0,\ldots,a_{11}$ are chosen such that $\ell_{\mbox{\scriptsize int}}(\beta)$ has the same
first $12$ Taylor coefficients as $\log Z(\beta)$.
The interpolated specific heat $c_{\mbox{\scriptsize int}}(\beta)$
is then determined according to (\ref{CL2b}).
It may happen that the ansatz (\ref{AC1}) exhibits poles in the physical temperature domain.
Then one can modify the largest exponents (i.e. $n=6$ in the enumerator  and $n=5$ in the
denominator) to find an appropriate ansatz without poles.

In what follows we will present the temperature
dependence of the specific heat  using a renormalized temperature $T/s(s+1)$.
This choice of the temperature scale enables the direct comparability of the
$c(T)$ profiles for different values of $s$ (cf. Ref.~\onlinecite{HTE_wir}).
First we consider $s=1/2$ and compare two interpolation results for the specific heat
with the exact result and also with `raw' HTE data and the [5,6]
Pad\'e approximant of the HTE series.
From the known complete set of eigenvalues we are also able to derive the
HTE series of the specific heat up to arbitrary orders (i.e. without using the
HTE code of Ref.~\onlinecite{HTE_wir}).
The results are presented in Fig.~\ref{figC1}.
If we only use the ``minimal" LTA data consisting of the ground state energy and the first excitation
(i.e. $k=1$ in Eq.~(\ref{AC1}))
and an HTE series up to order $n=11$
we obtain a specific heat curve that qualitatively
reproduces the two maxima, but gives a height of the first maximum that is $14\%$ too
low,  see Fig.~\ref{figC1}. Moreover, the position of the first maximum is slightly below the exact
position.
As expected, this result  is significantly better than
the  best Pad\'e approximant, that only reproduces the broad maximum at
higher temperature.

On the other hand one may ask which ``maximal data" for the LTA and HTE would give an optimal interpolation result. This is not
trivial since a very large order $n$ would produce badly conditioned matrices in the calculation of the Pad\'e coefficients. We found an optimum
by considering an HTE order of $n=35$ and taking into account the first $4$ excitations $\delta_i$ together with their degeneracies. Since $\delta_3>2\,\delta_1$
and $\delta_4>3\,\delta_1$, the simple form (\ref{CE3c}) of the LTA is no longer valid and has to be replaced by a variant involving higher terms of the Taylor series
of $\log(1+x)$.
The result of the ``optimal" interpolation then fits the exact
curve very well, see the red line of Fig.~\ref{figC1}, and has a maximal deviation of $4\%$ at the first maximum.

\begin{figure}
\begin{center}
\includegraphics[clip=on,width=70mm,angle=0]{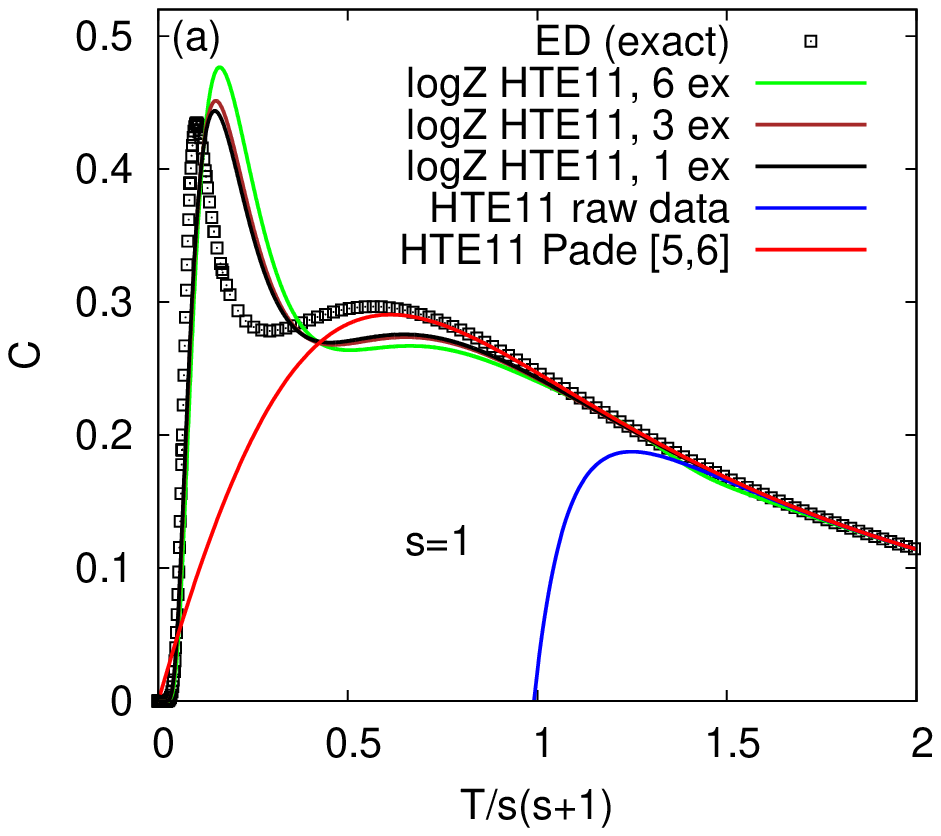}
\includegraphics[clip=on,width=70mm,angle=0]{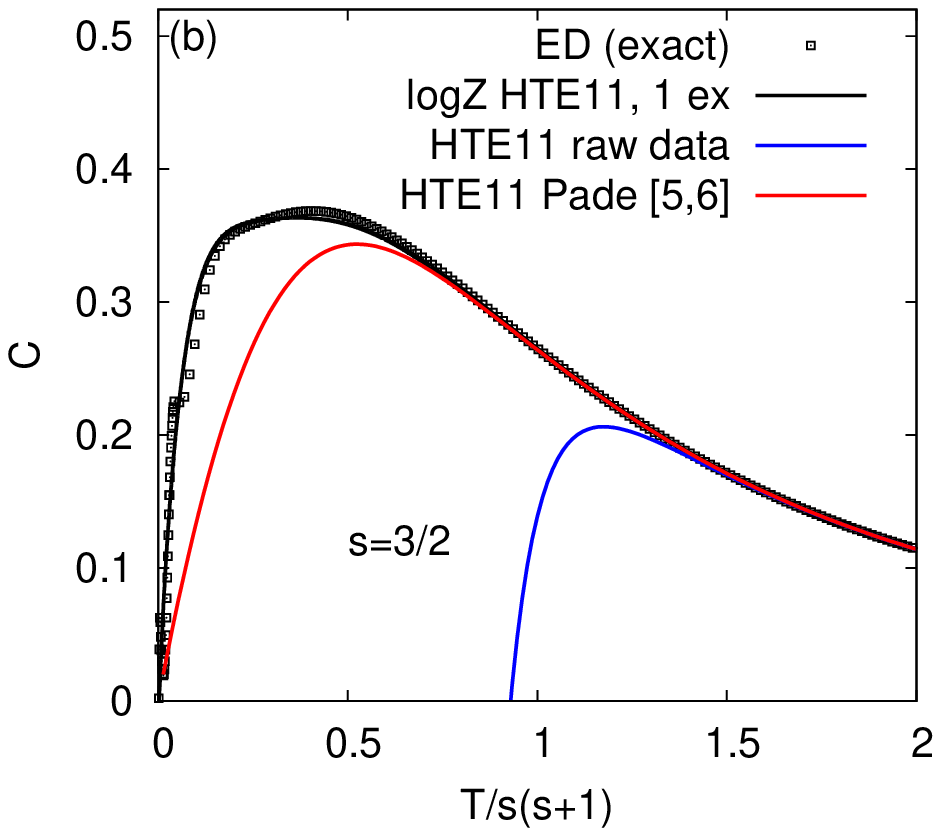}
\vspace*{1.0cm}
\end{center}
\caption{
Specific heat $c(T)$ of the cuboctahedron for $s=1$ (a) and $s=3/2$ (b), calculated by the Log Z method compared to the exact function
(ED, open squares).
The black solid line shows the  result using one excitation and an HTE of order
$n=11$.
For comparison we also show the `raw' HTE of order $n=11$ (blue) as well
as the [5,6] Pad\'e approximant (red). For $s=1$ we also show
 results using three (brown curve) and six excitations (green) and an HTE of order
$n=11$.
} \label{fig6}
\end{figure}
\begin{figure}
\begin{center}
\includegraphics[clip=on,width=70mm,angle=0]{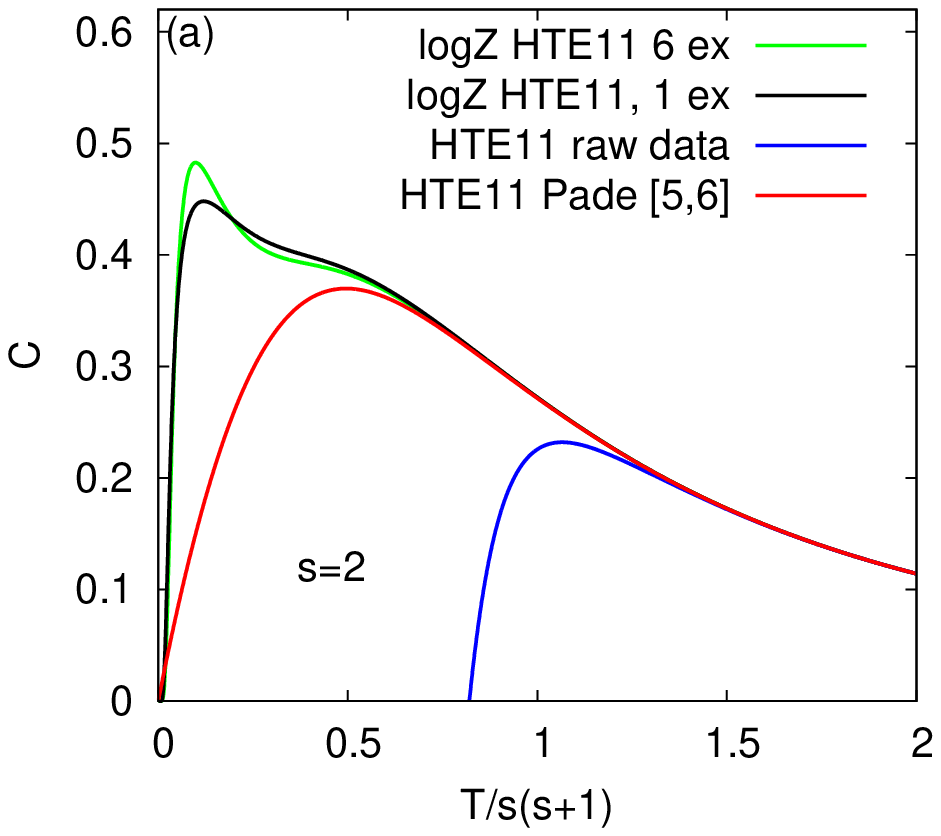}
\includegraphics[clip=on,width=70mm,angle=0]{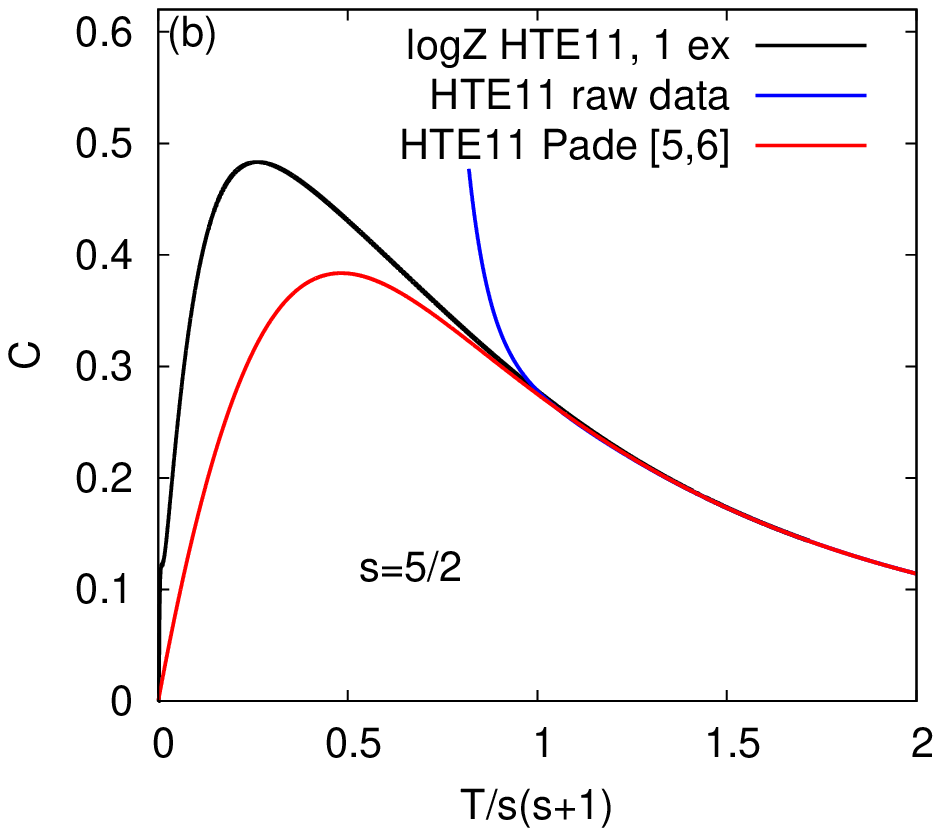}
\vspace*{1.0cm}
\end{center}
\caption{
Specific heat $c(T)$ of the cuboctahedron for $s=2$ (a) and $s=5/2$ (b), calculated by the Log Z method.
The black solid line shows the  result using one excitation and an HTE of order
$n=11$.
For comparison we also show the  `raw' HTE of order $n=11$ (blue) as well
as the [5,6] Pad\'e approximant (red). For $s=2$ we also show the log Z
interpolation using six excitations and an HTE of order
$n=11$ (green).
} \label{fig7}
\end{figure}

In  Fig.~\ref{fig6} we present the results
of the log Z
interpolation results and an HTE of order $n=11$  for spin quantum
numbers   $s=1$ and $s=3/2$.
Again we compare with the exact results and also with `raw' HTE data and
the [5,6] Pad\'e approximant of the HTE series.
The specific heat for $s=1$ also exhibits an extra low-temperature maximum
(located at $T/s(s+1)=0.101$) below the ordinary broad maximum, see
Fig.~\ref{fig6}a.
However, both maxima are not separated by a pronounced
minimum in $C(T)$ as it was
found for $s=1/2$ (see Fig.~\ref{figC1}). Rather they are smoothly
connected.
The important finding is that again the interpolation reproduces the extra
maximum.
Obviously,
the quantitative agreement is less good compared to the $s=1/2$ case.
Taking into account more than one excitation does not lead to an
improvement of the interpolation.
We conjecture that this is a hint to a general ``principle of balance" saying
that for a successful  interpolation the input of the HTE and the LTA should be balanced.
In practice, one will try to use the maximal number of HTE orders available,
as long as no poles appear in the interpolated functions.
It could be misleading to also use as much excitations and degeneracies as possible
that are, for example, obtainable by  Lanczos exact diagonalization.
This need not enhance the quality of the interpolation, but may even degrade it.
Using more excitations would
only improve the interpolation if simultaneously higher orders of the HTE  would be taken into account
as, e.~g., in the above case of $s=1/2$. Thus, for an HTE of order $n=11$ an LTA including, e.~g., one excitation
seems to be well balanced.\\
Moreover, we emphasize that interpolation with constraints is crucial for the correct
description of $c(T)$ for intermediate temperatures:
Both, the `raw' HTE and the [5,6] Pad\'e approximant do not give indications for
a second maximum. It may be argued that this second maximum is produced by the necessity for the interpolated
function $c_{int}(T)$ to allow for the two integral constraints (\ref{C1a}) and (\ref{C1b}).

The behavior of the specific heat for $s=3/2$ is somehow different, since
there is no pronounced low-temperature maximum, rather
$c(T)$ exhibits a broad plateau-like maximum in the region $0.13 \lesssim T/s(s+1) \lesssim 0.53$ (see
Fig.~\ref{fig6}b).
This feature is very well described by our interpolation scheme taking into
account the first excitation, only. On the other hand,  the `raw' HTE and the [5,6] Pad\'e approximant
fail to yield  a good description of this broad maximum.
At very low temperatures a tiny extra maximum is visible
in the exact data not seen in the interpolation.

Let us also emphasize that obviously the position of the broad maximum at moderate
temperatures is shifted to lower values of $T/s(s+1)$ as increasing
$s$. We find $T/s(s+1)=0.937$ ($s=1/2$), $T/s(s+1)=0.65$ ($s=1$) and $T/s(s+1)=0.406$
($s=3/2$).  Thus, we may speculate that the very existence of the double
maximum profile of $c(T)$ is a quantum effect and will disappear  as further increasing of
$s$.

From the above presented comparison of exact data and results of the log Z
interpolation it is evident that the minimal version, taking into account the
energies of the ground state and the first excitation, leads to very good
results for the specific heat in the whole temperature range. In particular,
specific features such as an extra maximum in the low-temperature region can
be reproduced by the interpolation scheme.
Thus we may conclude, that this scheme has some predictive power that can be
used for the investigation of systems, where no information on the full
spectrum is available, but the energies of the ground state and the  first excitation
are known.

In the next step we therefore apply our approach to $s=2,5/2,3$ and $7/2$.
Let us start with $s=2$ (i.e. first value of $s$ without exact data for
$c(T)$)
and $s=5/2$.
Interestingly, for $s=2$ (Fig.~\ref{fig7}a) a low-temperature maximum is found at
$T/s(s+1)=0.12$, but there is only a remnant of the ordinary broad maximum in form of
a plateau-like shoulder around $T/s(s+1) \sim 0.4$. Thus the general shape
is closer to that for $s=1/2$ and $s=1$ than that for $s=3/2$.

We may interpreted this observation as another indication
of the qualitative difference between half-integer and integer $s$ stemming from the fact that the
three spins on a triangle can be composed to a zero
total spin for integer $s$, whereas for half-integer $s$ the composed spin
on a triangle is
non-zero \cite{schmidt2005}.
This difference is also manifested in the ground state spin-spin correlation of the cuboctahedron
\cite{schmidt2005}.
This observation fits also  to the case $s=5/2$ shown in
Fig.~\ref{fig7}b, where $c(T)$ exhibits only one maximum without a shoulder.
 Concerning the low-energy spectrum, relevant for the low-temperature
behavior of $c(T)$,  we found for the excitation gap
$\Delta= 0.1165,
0.6705,
0.0434,
0.6234,
0.1006,
0.5237$, and
$0.12823$ for $s=1/2, 1, 3/2, 2, 5/2, 3$ and $7/2$, respectively, i.e. the
gap is significantly smaller for  half-integer $s$ than for integer $s$.

\begin{figure}
\begin{center}
\includegraphics[clip=on,width=70mm,angle=0]{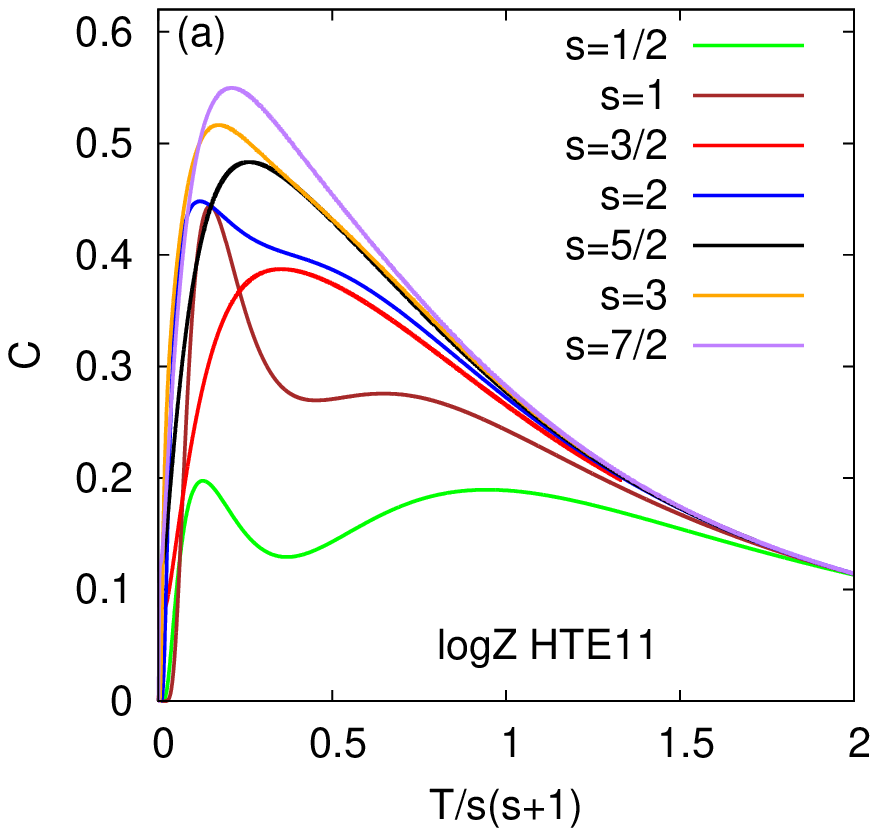}
\includegraphics[clip=on,width=70mm,angle=0]{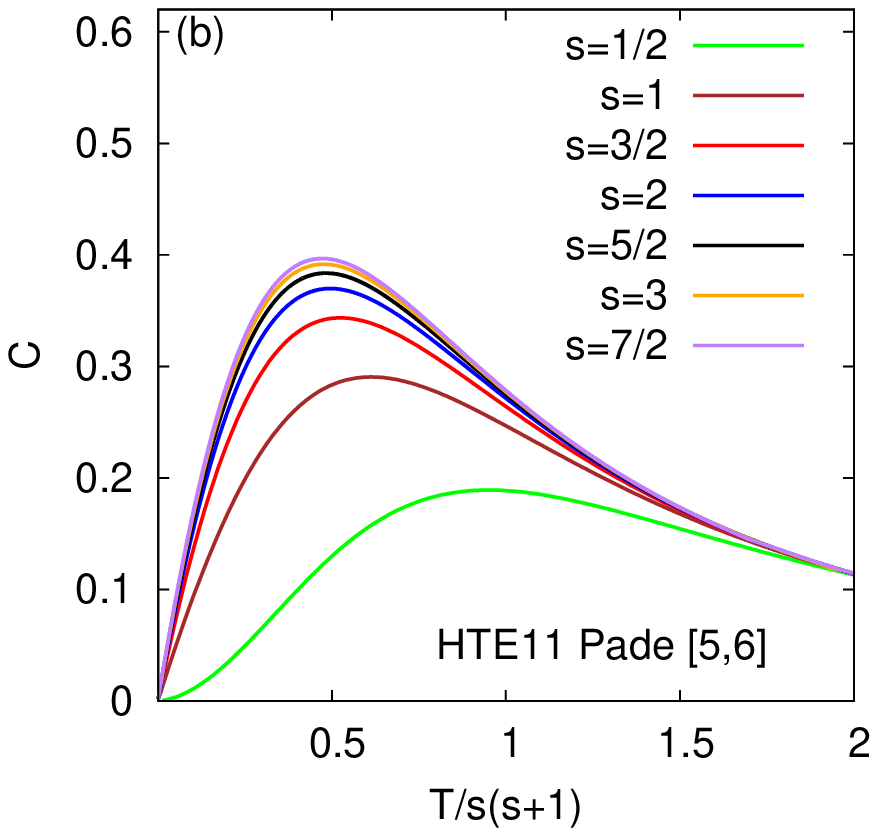}
\vspace*{1.0cm}
\end{center}
\caption{Specific heat $c(T)$ of the cuboctahedron for various values of the spin
quantum number $s$:  Comparison of the Log Z
interpolation (a) with the [5,6] Pad\'e approximants of the HTE
series of order $n=11$ without interpolation (b).
} \label{fig8}
\end{figure}

Let us summarize our findings for the spin-$s$ Heisenberg model on the
cuboctahedron. For that  we collect our log Z interpolation data  for all accessible values of $s$ in
Fig.~\ref{fig8}a.
For comparison we also show the [5,6] Pad\'e approximants in
Fig.~\ref{fig8}b.
Obviously, for $T/s(s+1) \gtrsim 1.8$ all curves
coincide, i.e. in  the renormalized temperature  scale the high-temperature
behavior is independent of $s$ \cite{HTE_wir}.
For  $T/s(s+1) \lesssim 1$
there is a strong influence of the spin $s$, where for low spin values $s$
some prominent extra  features  (shoulder, additional maximum)
emerge.
For larger spin $s \ge 5/2$ no significant extra features
appear, rather there is one pronounced maximum.
The  Pad\'e approximants describe the behavior of $c(T)$ reasonably well
down
to about  $T/s(s+1) \sim 0.6$, i.e. the extra features at low $T$ are not
covered by the  Pad\'e approximants.
The height $c_{\rm max}$ and the position $T_{\rm max}/s(s+1)$ of the maxima in $c(T)$
strongly depend on $s$. As an  overall tendency we observe that  $c_{\rm max}$
increases and  $T_{\rm max}/s(s+1)$ decreases with growing $s$,
however, there is not a simple  monotonic dependence on $s$ (compare $T_{\rm max}/s(s+1)$
for $s=5/2, 3$ and $7/2$ in Fig.~\ref{fig8}a).

\section{Interpolation for classical spin systems}\label{sec:CLA}

In the classical limit $s\rightarrow\infty$ the foundations of
interpolation described in section \ref{sec:C} have to be slightly reformulated.
This is already clear from the divergence of the r.~h.~s.~of the entropy integral (\ref{C1b}).
We will confine ourselves to the Log Z method.

Since the partition function $Z(\beta)$ diverges in the classical limit it has to be replaced
by the normalized partition function
\begin{equation}\label{CLA1}
\tilde{Z}(\beta)=\frac{\mbox{Tr }e^{-\beta H}}{\mbox{Tr }{\mathbbm 1}}=(2s+1)^{-N}\mbox{Tr }e^{-\beta H}
\;.
\end{equation}
The classical limit of (\ref{CLA1}) reads
\begin{equation}\label{CLA2}
Z^{\scriptsize\mbox{cl}}(\beta)=(4\pi)^{-N}\int_X e^{-\beta H(x)}\,dx
\;,
\end{equation}
where $X=\left({\mathcal S}^2\right)^N$ denotes the classical phase space of a system of $N$ spins
and $dx=d\Omega_1\ldots d\Omega_N$ its volume form. Consequently, the HTE
of $Z^{\scriptsize\mbox{cl}}(\beta)$
assumes the form
\begin{equation}\label{CLA3}
Z^{\scriptsize\mbox{cl}}(\beta)=\sum_{n=0}^\infty \frac{m_n}{n!}(-\beta)^n= 1+\frac{m_2}{2} \beta^2+\ldots
\;,
\end{equation}
where $m_n$ denotes the $n$ th normalized moment of $H$
\begin{equation}\label{CLA4}
m_n= (4\pi)^{-N}\int_X H^n(x)\,dx
\;,
\end{equation}
and $m_1=0$ has been used. From this one can derive the HTE of other thermodynamical functions
with leading terms
\begin{eqnarray}\label{CLA5a}
\ell(\beta)&\equiv& \log Z^{\scriptsize\mbox{cl}}(\beta)=\frac{m_2}{2}\beta^2+\ldots\quad,\\
\label{CLA5b}
c(\beta)&=& \beta^2 \frac{\partial^2}{\partial\beta^2}\ell(\beta)=m_2\beta^2+\ldots\quad,\\
\label{CLA5c}
s(\beta)&=& \ell(\beta)-\beta \frac{\partial}{\partial\beta}\ell(\beta)=-\frac{m_2}{2}\beta^2+\ldots\quad.
\end{eqnarray}
Regarding the low temperature asymptotic (LTA) of the classical partition function we assume the following
\begin{equation}\label{CLA6}
Z^{\scriptsize\mbox{cl}}(\beta)\sim e^{-\beta e_0}\,\beta^{-a}\,b\quad\mbox{for }\beta\rightarrow\infty
\;.
\end{equation}
Here $e_0$ denotes the ground state energy and $a>0,\,b$ are certain parameters.
(\ref{CLA6}) is, at least, satisfied
for certain finite classical spin system that we have investigated, see below. Since the purpose of our paper
is to demonstrate the applicability of certain concepts of interpolation and not to give a complete survey,
we will confine ourselves to the above case (\ref{CLA6}). It implies the LTA
\begin{eqnarray}\label{CLA7a}
\ell(\beta)&\sim&-\beta e_0 - a \log \beta + \ell_0,\\
\label{CLA7b}
c(\beta)&\sim& a,\\
\label{CLA7c}
s(\beta)&\sim& a+\ell_0-a \log \beta
\;,
\end{eqnarray}
where $\ell_0 \equiv \log b$. Hence $a$ can be identified with the finite limit
of the specific heat for zero temperature. (\ref{CLA7c}) implies that $s(T)$ diverges
for $T\rightarrow 0$ and hence the integral constraint (\ref{C1b}) has to be reformulated
for classical spin systems. We write
\begin{eqnarray}\label{CLA8a}
\Delta s &\equiv& s\left(\infty\right)-s\left(\frac{1}{\beta}\right) = \int_{\frac{1}{\beta}}^\infty \hspace{-1.5mm}\frac{d s}{d T}dT
= \int_{\frac{1}{\beta}}^\infty \hspace{-1.5mm}\frac{c}{ T}dT
\\
\label{CLA8b}
&=& \left[c(T) \log T \right]_{\frac{1}{\beta}}^\infty- \int_{\frac{1}{\beta}}^\infty \frac{d c}{d T}\log T \,dT
\;,
\end{eqnarray}
where (\ref{CLA8b}) is obtained by partial integration.
Since $c(T) \log T \sim m_2 T^{-2} \log T \rightarrow 0$ for $T\rightarrow \infty$, the first term in (\ref{CLA8b})
assumes the form $-c(\beta) \log \frac{1}{\beta}\sim a \log \beta$ for $\beta\rightarrow \infty$.
On the other hand, $s\left(\infty\right)-s\left(\frac{1}{\beta}\right) \sim 0 -a - \ell_0 +a \log \beta$
for $\beta\rightarrow \infty$ and thus
\begin{equation}\label{CLA9}
\int_{\frac{1}{\beta}}^\infty \frac{d c}{d T}\log T \,dT \sim a+\ell_0 \quad \mbox{for } \beta\rightarrow\infty
\;.
\end{equation}
The limit $\beta\rightarrow\infty$ of (\ref{CLA9}) yields the modified integral constraint
\begin{equation}\label{CLA10}
\int_{0}^\infty \frac{d c}{d T}\log T \,dT = a+\ell_0= a+\log\,b
\;,
\end{equation}
that holds for classical spin systems satisfying (\ref{CLA6}).
Further partial integrations yield an infinite number of integral constraints involving higher derivatives
of $c(T)$ that are, however, equivalent to (\ref{CLA10}).
In the following subsections we will test the considerations of this section and the Log Z interpolation for
an example that can be analytically solved.

\subsection{The equilateral triangle}\label{sec:CLAT}

The partition function of the equilateral spin triangle with Hamiltonian
\begin{equation}\label{CLAT1}
H={\mathbf s}_1\cdot {\mathbf s}_2 +{\mathbf s}_2\cdot {\mathbf s}_3 +{\mathbf s}_3\cdot {\mathbf s}_1
\end{equation}
can be calculated analytically. The result for $Z^{\scriptsize\mbox{cl}}(\beta,B)$
given in  Eq.~(14) of \cite{CLAL99}, where $B$ denotes the external magnetic field, can be evaluated in the limit $B\rightarrow 0$
and yields
\begin{equation}\label{CLAT2}
Z^{\scriptsize\mbox{cl}}(\beta)=\frac{\sqrt{\frac{\pi }{2}} e^{3 \beta /2} \left(-3\,
   \text{erfc}\left(\sqrt{\frac{\beta}{2}}\right)+\text{erfc}\left(3\,\sqrt{\frac{\beta}{2}}\right)+2\right)}{4\, \beta ^{3/2}}
\;.
\end{equation}
From this the specific heat can be derived in analytical form; but the result is too complicated to be
presented here.
The LTA of (\ref{CLAT2}) is easily seen to be of the form (\ref{CLA6}) with
\begin{equation}\label{CLAT3}
e_0=-\frac{3}{2},\quad a=\frac{3}{2},\quad b=\frac{\sqrt{\pi}}{2\sqrt{2}}
\;.
\end{equation}
This result can also be obtained without using the analytical form (\ref{CLAT2}) of the partition function, see appendix \ref{sec:LTA}.

\begin{figure}
\begin{center}
\includegraphics[clip=on,width=70mm,angle=0]{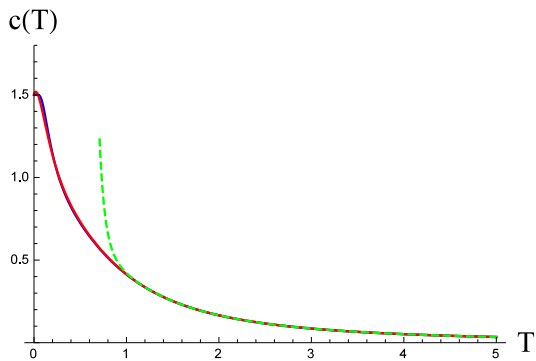}
\end{center}
\caption{The specific heat $c(T)$ of the classical equilateral spin triangle.
We show the analytical result (blue curve), the Log Z interpolation (red curve),
and the `raw' HTE of $12$th order (green dashed curve).
} \label{figPC}
\end{figure}

\begin{figure}
\begin{center}
\includegraphics[clip=on,width=70mm,angle=0]{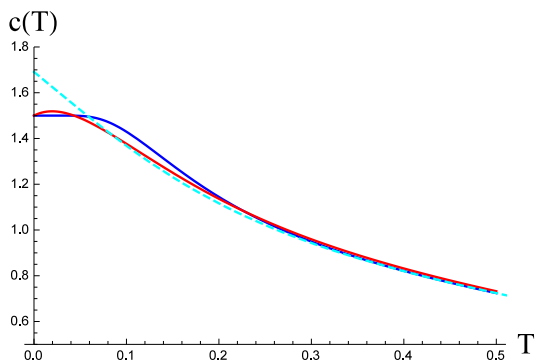}
\end{center}
\caption{The same specific heat as in Fig.~\ref{figPC}, but for lower temperatures.
We show the analytical result (blue curve), the Log Z interpolation (red curve),
and the HTE  [6,6] Pad\'e approximant (cyan dashed curve).
} \label{figPL}
\end{figure}

Next we consider interpolation of $c(T)$ according to the Log Z method. The LTA of $\ell(\beta)$ is not analytic at $\beta=0$,
see (\ref{CLA7a}). As a way out we consider the auxiliary function
\begin{equation}\label{CLAT7}
G(\beta)=\left(Z^{\scriptsize\mbox{cl}}(\beta)\right)^{2/3}
\;,
\end{equation}
that has the LTA
\begin{equation}\label{CLAT8}
G(\beta)\sim e^{\beta} \beta^{-1} \frac{\pi^{1/3}}{2}
\;.
\end{equation}
The factor $\beta^{-1}$ is still not analytical at $\beta=0$ but
can be compensated by a suitable choice of the Pad\'e function. It turns out that an optimal
interpolation is obtained by the ansatz
\begin{equation}\label{CLAT9}
G_{\scriptsize\mbox{int}}(\beta)= \frac{\pi^{1/3}  e^{\beta } \left(a_4 \beta ^4+a_3 \beta ^3+a_2
   \beta ^2+\frac{2 \beta }{\pi^{1/3}}\right)}{2 \beta  \left(a_4
   \beta ^4+a_7 \beta ^3+a_6 \beta ^2+a_5 \beta +1\right)}
\;.
\end{equation}
The Pad\'e coefficients $a_2,\ldots, a_7$ are, as usual, determined by the condition that $G_{\scriptsize\mbox{int}}(\beta)$
has the same first six HTE coefficients as $\left(Z^{\scriptsize\mbox{cl}}(\beta)\right)^{2/3}$. The latter
can be calculated from the analytical form of $Z^{\scriptsize\mbox{cl}}(\beta)$, but also independently by using
the $C^{++}$ program provided at \cite{url_HTE10}.
At the end, the interpolation of $G(\beta)$ thus obtained has to be transformed
into interpolations of $Z^{\scriptsize\mbox{cl}}(\beta)$ and $c(T)$.
We compare this result with the analytical form of the specific heat and
with approximations based solely on HTE data of $12$-th order,
see the Figs.~\ref{figPC} and \ref{figPL}.
One observes that in this case the HTE  [6,6] Pad\'e approximant of $c(T)$ assumes the finite value
of $c(0)=1.69087\ldots$ that is, however, $13\%$ above the correct value  $c(0)=\frac{3}{2}$.
We have also checked the integral constraints (\ref{C1a}) and (\ref{CLA10})
for the analytical and the interpolation form for the specific heat by numerical integrations.
The close coincidence between analytical and interpolation results
shows the consistency of the present method.

\section{Summary}
In our paper we present an approach to evaluate the specific heat $c(T)$ of
magnetic systems using an interpolation scheme between the
known low-temperature and high-temperature properties of $c(T)$ that also
exploits sum
rules constraining the specific heat. To satisfy these sum
rules for $c(T)$ it is more convenient
to perform the interpolation
for the logarithm of the partition
function $Z$ (i.e. the free energy $F/T$).
The requested input at high temperatures in form of a high-temperature expansion
series of log $Z$ can
be  obtained
by a simple C++-program \cite{,HTE_wir,url_HTE10}. The input at
low temperatures in form of the ground-state energy and the behavior of
$c(T)$ as $T\to0$ can be provided by the toolbox of many-body methods designed for the
low-energy degrees of freedom.
As a result,
the proposed interpolation scheme represents a quite universal and powerful
instrument to study the specific heat, e.g., for frustrated quantum magnets and to
provide model data to compare with experimental results.
We demonstrate the accuracy of our approach
by comparing the approximate interpolation data with
exact numerical date for a nontrivial strongly frustrated model system, the
spin-$s$ Heisenberg antiferromagnet on the cuboctahedron.
In particular, we found evidence that a prominent feature in form of
an additional low-temperature maximum in $c(T)$ can be detected by the
interpolation approximation.
We may conclude that the log Z interpolation scheme has some predictive power that can be
used for the investigations of strongly frustrated quantum spin systems, where
other tools, such as the quantum Monte Carlo technique, are not applicable.

\appendix
\section{LTA of the classical spin triangle}\label{sec:LTA}

We will show how to obtain the result (\ref{CLAT3}) that is needed for the interpolation of the specific heat without
using the analytical form of $Z^{\scriptsize\mbox{cl}}(\beta)$. In fact, the LTA of $Z^{\scriptsize\mbox{cl}}(\beta)$
can be determined by the $3$-dimensional Laplace method, see \cite{F11}.
First, it is clear that the lowest energy of (\ref{CLAT1})
is realized by any coplanar state of three unit vectors forming mutual angles of $\frac{2\pi}{3}$ and only by such states.
Hence the lowest energy is $e_0=-\frac{3}{2}$ and the usual rotational degeneracy of the ground state is the only one.
This can be further supported by calculating the eigenvalues of the Hessian of (\ref{CLAT1}). When doing this one has to be
careful in choosing the right coordinates. We fix the vector
\begin{equation}\label{LTA1}
{\mathbf s}_1=\left(\begin{array}{c}0\\0\\1\end{array}\right)\;,
\end{equation}
and write
\begin{eqnarray}\label{LTA2a}
   {\mathbf s}_2&=&\left(\begin{array}{c}\sqrt{1-(-1/2+z_2)^2}\\0\\-1/2+z_2\end{array}\right)\;, \\
\label{LTA2b}
  {\mathbf s}_3&=&\left(\begin{array}{c}\sqrt{1-(-1/2+z_3)^2}\cos (\pi+\varphi_3)\\\sqrt{1-(-1/2+z_3)^2}\sin (\pi+\varphi_3)\\-1/2+z_3
  \end{array}\right) \;,
\end{eqnarray}
thereby utilizing full rotational symmetry.
Thus the reduced phase space (the full phase space reduced by the rotational group) is described by three canonical
coordinates ${\mathbf u}=(u_1,u_2,u_3)\equiv(z_2, z_3, \varphi_3)$ and the ground state corresponds to ${\mathbf u}={\mathbf 0}$.
We consider the matrix $H_2$ that is, up to a factor $1/2$, the Hessian of (\ref{CLAT1}), i.~e.~the symmetric $3\times 3$-matrix
formed by all second derivatives of the Hamiltonian $H$ w.~r.~t.~the $u_i,u_j$,
evaluated at the ground state such that
\begin{equation}\label{CLAT4}
H=e_0 + {\mathbf u}\cdot H_2 \cdot {\mathbf u} + {\mathcal O}(|{\mathbf u}|^3)
\;.
\end{equation}
The three eigenvalues of $H_2$ are $h_1=1,\;h_2=\frac{3}{8},\; h_3=\frac{1}{3}$. They are positive in accordance with the
fact that (\ref{CLAT4}) is the expansion of $H$ at the ground state. Let ${\mathbf v}=(v_1,v_2,v_3)$ denote the coordinates
corresponding to the eigenbasis of $H_2$ that are obtained by a suitable rotation of ${\mathbf u}$. For low temperatures
the system will stay close to the ground state and its energy $H$ can be well approximated by the second order Taylor series (\ref{CLAT4}).
For the integrand $e^{-\beta H}$ occurring in the integral defining the partition function we may perform the approximation
\begin{equation}\label{CLAT5}
e^{-\beta H}\approx e^{-\beta (e_0 + {\mathbf u}\cdot H_2 \cdot {\mathbf u})}= e^{-\beta \left(e_0 + \sum_{i=1}^3 h_i v_i^2\right)}
\;.
\end{equation}
The integral of (\ref{CLAT5}) over the reduced phase space can hence be approximated, besides the constant factor $e^{-\beta e_0}$,
by the product of three Gaussian integrals of the form
\begin{equation}\label{CLAT6}
\int_{-\infty}^{\infty} \exp\left(-\beta h_i \, v_i^2\right)\,d v_i= \sqrt{\frac{\pi}{\beta h_i}},\;i=1,2,3
\;.
\end{equation}
The product of the three integrals (\ref{CLAT6}) gives
$\frac{\pi^{3/2}}{\beta^{3/2} \sqrt{h_1 h_2 h_3}}=\frac{\pi^{3/2}\sqrt{8}}{\beta^{3/2} }$ and has further to be divided
by $8\pi$ due to normalization. Note that the missing factor $8\pi^2$ is the volume of the rotational group that has to be left out
since we integrate only over the reduced phase space. This gives the correct term $\beta^{-3/2}$, i.~e.~, $a=\frac{3}{2}$ in
(\ref{CLA6}), and, further, the factor $b=\frac{\pi^{3/2}\sqrt{8}}{8 \pi}=\sqrt{\frac{\pi}{8}}$ in accordance with (\ref{CLAT3}),
which completes the calculation of the LTA of the equilateral triangle without using the analytical result (\ref{CLAT2}).

\section*{Acknowlegedments}
We thank J\"urgen Schnack for providing the exact data for the specific heat
of the cuboctahedron with $s=3/2$.
Moreover, we are grateful to Gregoire Misguich for committing to us the unpublished HTE
coefficients up to order $n=20$ of the $s=1$ Haldane chain that were originally computed by Elstner, Jolicoeur, and Golinelli.


\end{document}